# Antimatter

Beatriz Gato-Rivera

Instituto de Física Fundamental, CSIC

Serrano 123, 28006 Madrid, Spain

In this article we concisely explain: what antimatter is, its differentiation between primordial and secondary, how it is produced, where it can be found, the experiments carried out at CERN to create and analyze antiatoms, the problem of the matter-antimatter asymmetry, and the medical and technological applications of antimatter in our society.

Antimatter, the reverse of matter, is one of the most fascinating aspects of particle physics and also one of the most unknown. Surprisingly, an object made of antimatter would be indistinguishable, judging from its appearance, from one made of matter. Indeed, if antimatter stars existed, they would shine identically to their matter counterparts, emitting the same light. Without knowing it, we live surrounded by antimatter and by the radiation resulting from its annihilation against matter. For example, on the Earth's surface we are subjected to an incessant rain of particles, both matter and antimatter particles, produced in the cosmic ray cascades.

Furthermore, it is estimated that 10% of the visible light that reaches us from the Sun is due to the matter-antimatter annihilation taking place in its interior, since the nuclear furnaces of stars are great producers of antielectrons that immediately annihilate with the electrons present in the plasma in which they are immersed. Moreover, in our society we make extensive use of antimatter, especially antielectrons, both in medicine - being the essential ingredient of PET scans - and in industry and research related to Materials Science and Technology.

## 1. What is antimatter?

To understand what antimatter is, one must enter the world of elementary particles and from there it can be said that antimatter is the reverse of matter because particles and their antiparticles have opposite properties, in addition to some identical ones that do not admit opposite values. Thus, particles and their antiparticles have the same mass, the same spin and the same half-life, but opposite values of the strong charge, weak charge, electric charge, baryon number, lepton number and helicity. All particles have antiparticles, although in some cases they coincide with each other, as happens with the photon and the Higgs boson.

The Standard Model that describes elementary particles is based on quantum field theory, which combines quantum mechanics with special relativity to describe particles at very high speeds. According to this theory, elementary particles are just excitations of quantum fields, which fill all of space, the entire Universe; and each species of particle has its own field, except that a particle and its antiparticle share the same field. It is not surprising, therefore, that particle-antiparticle pairs are often produced when quantum fields are excited by the energy of collisions, either in particle accelerators or in the atmosphere by the action of cosmic rays.

The first antiparticle found, the antielectron, was named *positron* by its discoverer, Carl Anderson, a young researcher at Caltech, California, who investigated the traces left by particles from the cosmic radiation in the small cloud chamber of his laboratory. The finding was published in 1932 and, interestingly, other scientists realized that they had also observed these curious traces, similar to those left by electrons, but spinning 'backwards', denoting positive electric charge. It so happens, moreover, that the existence of the antielectron was proposed only one year earlier, in 1931, by the theoretical physicist Paul Dirac, to account for the negative energy solutions of his famous equation describing electrons at relativistic speeds, published in 1928.

As for antimatter atoms, it is not known if they exist in nature somewhere in the Universe; the only ones that we have known have been created in laboratories, starting in 1995, and almost uniquely at CERN's Antimatter Factory.

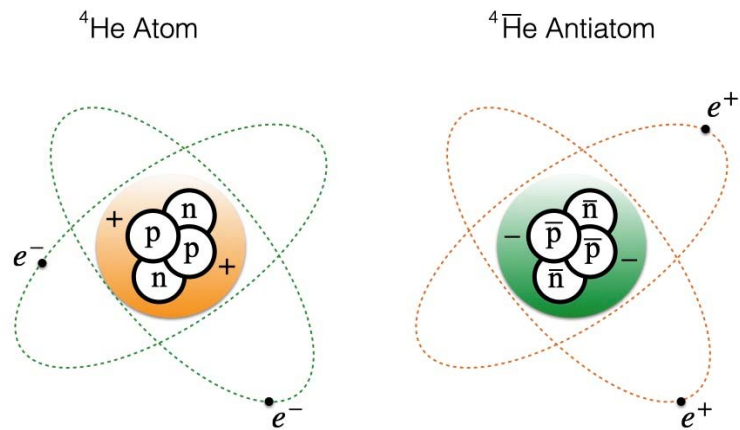

**Fig. 1.** Sketch of a Helium-4 atom and an Antihelium-4 antiatom. The nucleus of the atom, composed by two protons and two neutrons, has positive electric charge while the nucleus of the antiatom, composed by two antiprotons and two antineutrons, is negative. These nuclei are named alpha particles and antialpha particles, respectively.

2. Matter-antimatter annihilation

A very distinctive feature of antimatter is that when it comes into contact with matter, they annihilate each other. If the encounter between a particle and its antiparticle takes place at small velocities, i.e. at low energies, the outcome of the annihilation consists only of photons; but if the collision occurs at sufficiently high energies, other particles and antiparticles can also be created, in addition to photons. This phenomenon is a common occurrence in particle collisions, and is due to Einstein's formula for the conversion between mass and energy, $E = mc^2$, whereby part of the energy of the collision can be invested in creating particles and antiparticles, when exciting their quantum fields. Since photons are massless, no threshold energy is required for them to be produced.

It should also be noted that matter-antimatter annihilation is the most energetic process that exists. For example, a gram of antimatter dropped on our planet would produce a deflagration equivalent to almost three times the atomic bomb that devastated Hiroshima in 1945, assuming it had about 15 kilotons of TNT. Despite this, it is not possible to use matter-antimatter annihilation to cover the energy needs of our society as we do with nuclear energy; in fact, we are very far from being able to achieve it.

The fundamental reason is that with current technology it is only possible to produce derisory amounts of antimatter and furthermore at the cost of a gigantic technological, energetic and financial effort. For example, the largest antiproton factory that has ever existed, in charge of supplying them to the Tevatron accelerator of the Fermi National Laboratory (Fermilab, USA), managed to produce about $10^{15}$ antiprotons per year, barely enough to bring a liter of water at room temperature to the boiling point by annihilating them with the same number of protons (1.4 x $10^{15}$ $p\,\bar{p}$ annihilations would be necessary to raise the temperature of one liter of water from 0 ºC to 100 ºC). And if this were not enough, taking into account that one gram of antiprotons is equivalent to 6.02 x $10^{23}$ of them, it would take 602 million years to produce a single gram using the technology of the Fermilab factory. As for CERN, using its current technology it would take 60,200 million years to produce one gram of antiprotons, more than four times the 13,800 million years that the Universe has been in existence.

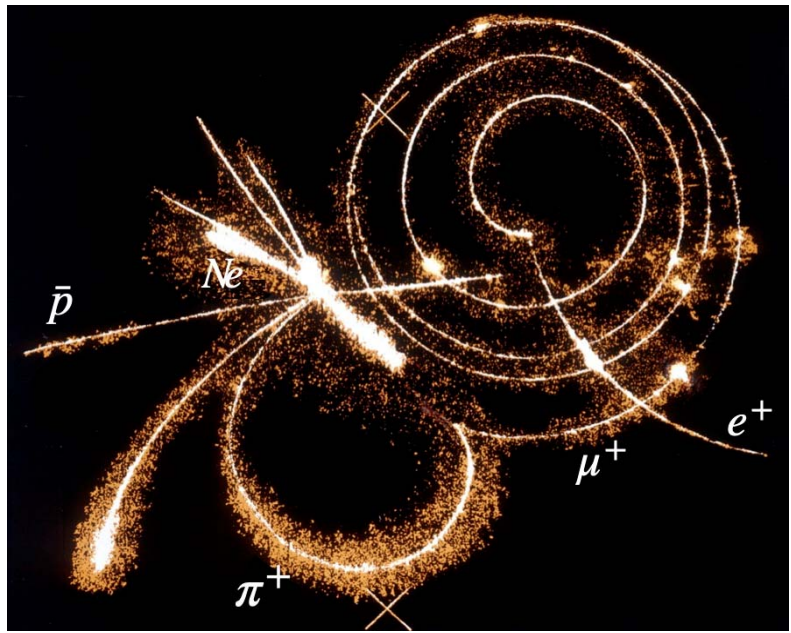

**Fig. 2.** Antiproton annihilation against a nucleus of a neon atom. Experiment PS-179 in the LEAR machine at CERN, in 1984. Among the products of the annihilation, a positive pion is created that moves in a spiral, due to the effect of the applied magnetic field, and decays producing an antimuon. This one revolves several times until it decays in turn, giving as a result a positron (antielectron). In each of these decays neutrinos and/or antineutrinos are also emitted, but they cannot be observed since being neutral they leave no trail. *Credit* Courtesy of CERN.

## 3. Sources of antimatter

It is crucial to differentiate between primordial and secondary antimatter. The former was created at the beginning of the Universe, in the first moments after the Big Bang, about 13,800 million years ago, and may well have disappeared almost entirely. Secondary antimatter, on the contrary, is continuously being created in our environment, either by collisions between particles, as happens in cosmic ray cascades, or in nuclear reactions within stars and other astrophysical processes, or else in the nuclei of certain radioactive substances, like the potassium isotope $^{40}K$, which is abundant in bananas, in walls and in our own bones. In fact, this isotope emits 89.28% of the time one electron plus one antineutrino, i.e. $\beta$ radiation; it emits 10.72% of the time gamma radiation plus one neutrino; and it emits 0.001% of the time one positron plus one neutrino, the so-called $\beta^+$ radiation.

Positrons emitted by the bones do not get out of the body, since they are annihilated by the electrons as they pass through, and their path is barely one millimeter. In the case of bananas, which also have $^{40}K$ in their skin, about 15 positrons escape to the outside every 24 hours. Hence bananas emit antimatter, and the same can be said of ceramics, walls and the like, since lime and cement also contain $^{40}K$. The positrons detected by Carl Anderson in his laboratory came from cosmic ray cascades and were produced by the decay of antimuons that passed through the ceiling.

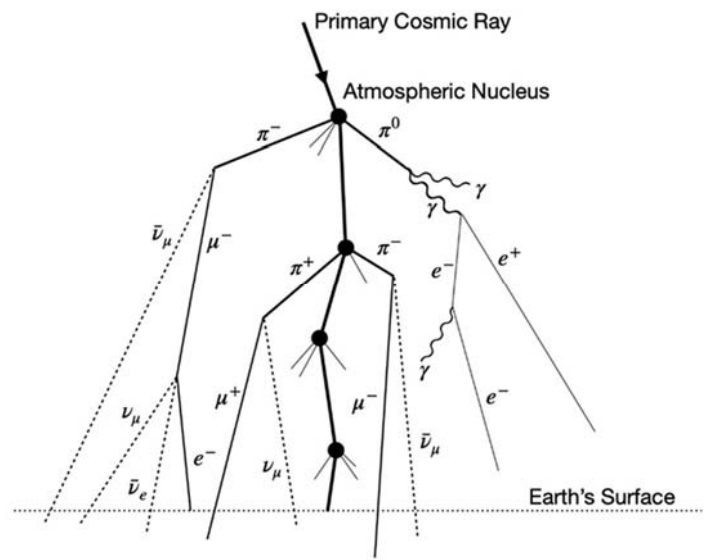

**Fig. 3.** Sketch of a cosmic ray cascade, where matter and antimatter particles are found in equal numbers, approximately.

The main sources of antimatter within our reach are particle accelerators and our own atmosphere, where cosmic rays collide with the atoms present there, producing large cascades of secondary radiation, the more energetic the primary rays, the larger. Surprisingly, cosmic rays can reach energies several orders of magnitude higher than the highest energies that can be attained in accelerators and, as we said, in both cases part of the energy produced in the collisions is invested in creating other particles, both matter and antimatter particles (approximately half and half).

Cosmic ray cascades can be made up of billions of particles, many of which reach the Earth's surface. For example, at sea level it is estimated that one muon (or antimuon) arrives every minute per $cm^2$, and there also arrive, between 10 and 100 times more, both electrons and positrons, as well as myriads of neutrinos, antineutrinos and photons. The most energetic muons and antimuons can even penetrate our homes and pass through several kilometers underground before disintegrating, while neutrinos and antineutrinos traverse the entire planet with hardly a flinch.

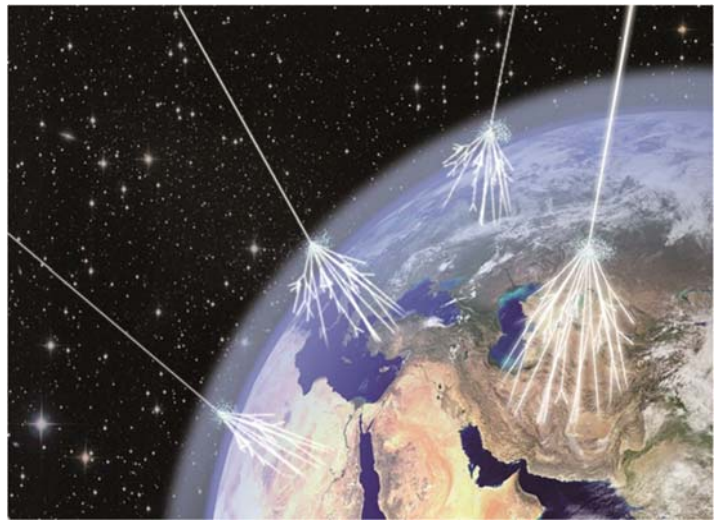

**Fig. 4.** Artistic recreation of primary cosmic rays impacting on Earth's atmosphere and giving rise to cascades of secondary radiation. *Credit* Asimmetrie/INFN.

On the other hand, the secondary radiation can also leave the atmosphere outwards and, in the case of electrically charged particles, they can remain trapped in the Earth's magnetic field as part of the Van Allen radiation belts. In this respect, it should be noted that in 2011 the PAMELA detector, on board the Russian satellite Resurs-DK1, found a sub-belt of antiprotons in the inner belt, produced when highly energetic cosmic rays collide with molecules from the upper layers of the atmosphere. The detector also found that the antiproton flux there exceeds that of the antiprotons traveling freely through space by a factor of 1,000, making this Van Allen sub-belt the most abundant source of antiprotons near the Earth.

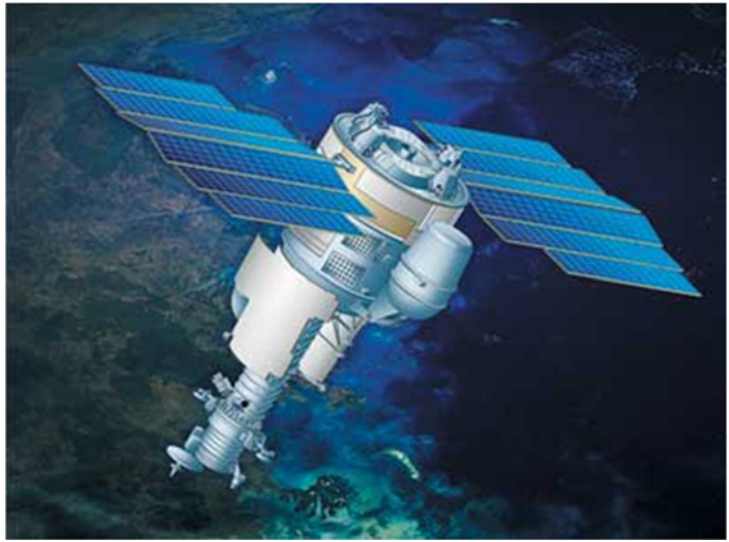

**Fig. 5.** The PAMELA cosmic ray detector, that looks like a dice cup, on board the Russian Resurs-DK1 satellite. Among its many achievements, in 2011 it discovered an antiproton sub-belt in the inner Van Allen belt. *Credit* Courtesy of the PAMELA collaboration.

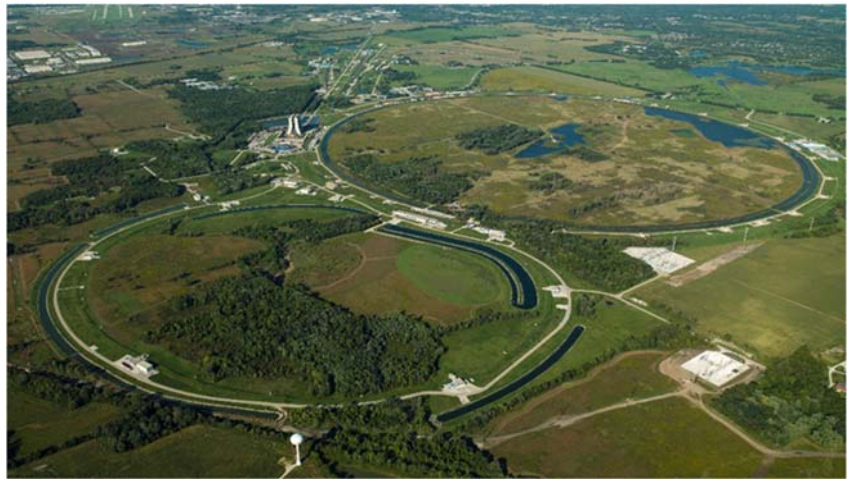

**Fig. 6.** Aerial view of the Fermi National Accelerator Laboratory (Fermilab), 60 km from Chicago. The ring in the background, 6.3 km long, was the Tevatron, a proton-antiproton collider that was the most powerful accelerator in the world from 1985 to 2009, and was shut down in 2011. *Credit* Courtesy of Fermilab.

In addition to nuclear reactions in stars, there are many other astrophysical processes that produce antimatter as well, and even release it into outer space, so that a small fraction of the primary cosmic rays, less than 1%, consists of antiprotons and positrons (98% are protons and alpha particles, and the remaining 2% consists mainly of atomic nuclei). Accordingly, since 2006 attempts are being made to capture atomic nuclei of antimatter larger than antiprotons in outer space, such as antialpha particles consisting of two antiprotons and two antineutrons. The first detector launched to this end was PAMELA, whose mission ended in 2016, and since 2011 the AMS-02 detector is operating, installed aboard the International Space Station. But so far this search has not been successful.

4. Experiments with antiatoms

In 1995 a team of scientists from CERN succeeded in producing the first antiatoms, a total of nine antihydrogen $\bar{H}$ atoms, made up of a single antiproton in the nucleus and a positron in the shell. Two years later, Fermilab announced that it had managed to produce another 100 antihydrogen atoms. In both cases, they were moving very fast and annihilated with the surrounding matter before they could be analyzed.

This situation changed in the year 2000, when the CERN Antiproton Decelerator (AD) came into operation, making it possible to produce slower $\bar{H}$ atoms. It was installed in a large room together with several experiments, to which it supplied antiprotons in order to create $\bar{H}$ atoms and analyze them. These experiments, with the names: ATRAP, ALPHA, ASACUSA and AEGIS, were discontinued in September 2018 due to a technical three-year shutdown for upgrading the large accelerator LHC. They resumed operation in 2021 along with the new GBAR experiment and the new ELENA decelerator, which is coupled to the AD decelerator and allows to reduce the speed of antiprotons 50 times more. To be precise, ELENA slows down the antiprotons coming from the AD decelerator with an energy of 5.3 MeV, to an energy of 0.1 MeV.

In the experiments performed so far, the researchers have utilized spectroscopy techniques mainly - either with laser or with microwaves - to check whether the energy levels of the $\bar{H}$ atoms are identical to those of the $H$ atoms, judging by the spectra of emission and absorption of photons. To date, no difference has been observed between them.

The ALPHA collaboration is leading these investigations. In December 2016, they published the first observation of a spectral line in antimatter atoms, the one corresponding to the 1S-2S transition in $\bar{H}$ atoms. The method followed consisted of irradiating the antiatoms, confined in a magnetic trap, with two opposing laser beams of ultraviolet light of wavelength λ = 243 nm - which is resonant with the 1S-2S transition in *H* atoms - because the absorption of two of these photons facing each other makes the electron jump from the 1S orbital to the 2S orbital. Thus, if the $\bar{H}$ atoms had exactly the same energy levels as the *H* atoms, this technique would cause that a large part of the positrons that were in their ground state in the 1S orbital would get excited and would jump to the 2S orbital, as actually happened.

Shortly afterwards, in 2018, the ALPHA collaboration published the first observation of the Lyman-α spectral line, with λ = 121.57 nm, in $\bar{H}$ atoms, corresponding to the 1S-2P transition; but a year earlier they had already succeeded in inducing hyperfine atomic transitions within the fundamental 1S level by exposing the $\bar{H}$ atoms to microwave radiation. These sublevels originate from the interaction between the spin of the positron and the spin of the antiproton in the atomic nucleus.

Other hyperfine transitions followed, within the 2P level, and with the results of all these observations the team was able to deduce for the $\bar{H}$ atoms the counterpart of the Lamb effect, which is the energy difference between the 2S and 2P levels with total angular momentum $j = 1/2$. These levels are separated by a tiny energy of 4.372 x $10^{-6}$ eV, and the effect is due to the interaction of the electrons, or positrons, with the quantum fluctuations of the electromagnetic field in the vacuum, as predicted by quantum electrodynamics. This result was published in *Nature*, in February 2020.

It should also be noted that the ASACUSA experiment creates and investigates antiprotonic helium, in which an electron of the helium atom is replaced by an antiproton, thus producing the so-called 'atomcules', hybrids composed of matter and antimatter, which behave like an atom and a molecule at the same time.

Curiously, there are also attempts to elucidate whether the $\bar{H}$ atoms fall in the Earth's gravitational field with the same acceleration as matter; that is, if antimatter participates in the universality of free fall. No results have been obtained yet, apart that antimatter indeed falls down, as most physicists believed (there were those who bet on the possibility that antimatter could feel gravitational repulsion towards matter, so that the $\bar{H}$ atoms, instead of falling, would rise in the Earth's gravitational field).

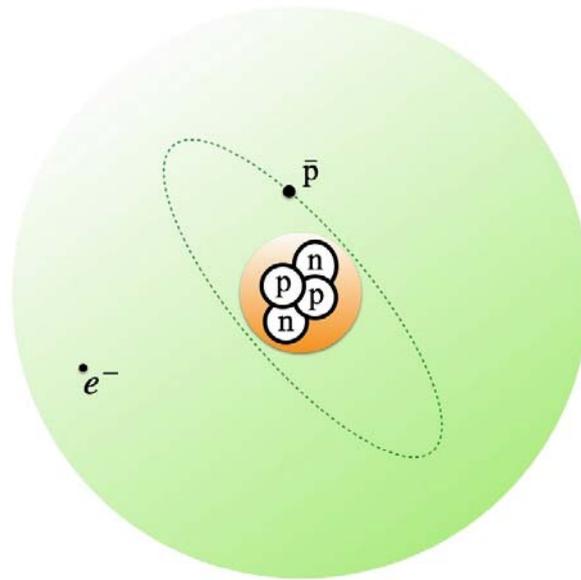

**Fig. 7.** Sketch of an 'atomcule' of antiprotonic helium. It is a hybrid between matter and antimatter particles and is halfway between an atom and a molecule since it is composed of two atomic nuclei, like a molecule, but their electronic structure is like that of an atom. Due to its bigger mass, the orbit of the antiproton around the helium nucleus is deep inside the electronic cloud occupied by the electron.

## 5. Primordial antimatter

The primordial antimatter could have been completely annihilated against matter at the beginning of the Universe; but if it exists today, it could have given rise to structures such as antistars, antiplanets, and even small antigalaxies. Antistars would be indistinguishable from ordinary stars judging by their appearance, since they would emit the same electromagnetic spectrum, but we could differentiate between them because hydrogen-consuming stars, like the Sun, emit neutrinos, while their antimatter counterparts would emit antineutrinos. However, since we do not have any instruments yet that allow us to elucidate whether a star emits neutrinos or antineutrinos, with the exception of the Sun, we will have to wait for future technologies that can do so.

Nevertheless, there are several types of astronomical observations that can give us information as to whether antimatter structures exist. These observations fall on three fronts: the cosmic background of diffuse gamma radiation, DGRB; the cosmic microwave background radiation, CMB, which was released when the first atoms were formed, some 380,000 years after the Big Bang; and cosmic rays in space.

DGRB and CMB radiations could reveal annihilation boundaries between matter domains and antimatter domains, which are not yet known. Of course, the impossibility of detecting such boundaries does not mean that this possibility should be ruled out, and the non-observations are useful to obtain estimates on the abundance of the primordial antimatter versus matter, which must be less than one antiproton among tens of millions of protons. Therefore, at present the possibility that there might exist small islands of antimatter throughout the Universe cannot be ruled out, even within our own galaxy, since they would escape detection.

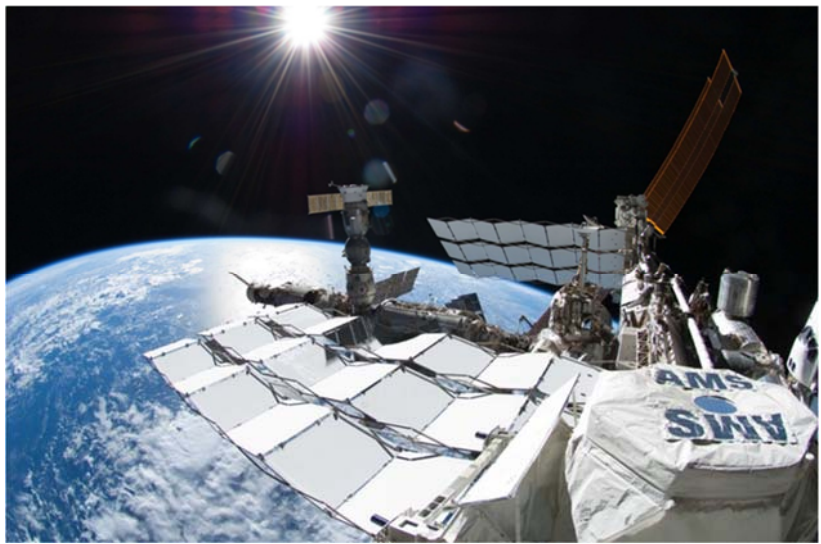

**Fig. 8.** The Alpha Magnetic Spectrometer (AMS-02) on board the International Space Station (ISS). Its primary purpose is to search for antimatter, as well as for signals of dark matter annihilation, although it also makes various types of measurements on cosmic rays with unprecedented precision. *Credit* Courtesy of NASA.

Cosmic rays, on the other hand, could provide irrefutable evidence for the existence of antistars if anticarbon nuclei, or larger ones, were discovered, which could only have been forged in nuclear reactions within them, just as carbon nuclei could only have been created from nuclear reactions in ordinary stars. The detection of antihelium nuclei, however, would not even prove that primordial antimatter exists, because these nuclei could have been formed in very energetic ordinary astrophysical processes. At least this is what many experts think, although there is some discrepancy among them.

## 6. Matter-antimatter asymmetry

The problem of the matter-antimatter asymmetry in the Universe stems from the assumption that particles and their antiparticles were created in identical amounts in the first moments after the Big Bang (or if they were not identical, cosmological inflation took care of making them so). For if they were created in equal numbers in a tiny, extremely dense and hot Universe, they should have annihilated each other almost entirely. So, how did matter survive that Great Annihilation that practically wiped out antimatter? Actually, the Great Annihilation almost put an end to matter as well, because, one second after the Big Bang, for every primordial proton that survived billions succumbed to extinction, and that proton ended up immersed in a bath of billions of photons that, 380,000 years later, gave rise to the microwave background radiation CMB that we observe today. For these reasons, something is supposed to have happened, before the Great Annihilation, in order to generate the very slight surplus of particles over antiparticles that was sufficient for the material Universe to come into existence, as we know it.

As a matter of fact, particles and their antiparticles do not always behave identically (apart from the fact that they have opposite properties). For example, in some cases they disintegrate at different rates in specific decay channels. This happens with the tau lepton, the $K^0$ mesons and all B-type mesons (mesons are made up of a quark and an antiquark). This phenomenon is known as violation of the CP symmetry, where C stands for charge conjugation and P for parity. This symmetry exchanges particles and antiparticles with each other, and its violation constitutes one of the three conditions that Andrei Sakharov proposed so that processes between particles can generate a surplus of matter over antimatter.

The other two Sakharov conditions are the violation of the baryon number B and that the processes occur outside thermal equilibrium; i.e. that they are irreversible. These three conditions were fulfilled during the primordial expansion of the Universe, but they could not possibly produce the amounts of matter observed in the Universe - but rather quantities several orders of magnitude smaller - as one deduces by analyzing them in the light of the Standard Model. Consequently, the creation of primordial protons, neutrons and electrons (baryogenesis and leptogenesis), remains one of the most challenging problems in particle physics at the present time.

## 7. Medical and technological applications of antimatter

The annihilation of electrons from biological matter with positrons provides one of the most important imaging techniques used in hospitals. This is Positron Emission Tomography (PET), in which substances containing short-lived positron emitting isotopes are injected into the patient. These positrons annihilate with electrons from the biological tissue and emit two 511 keV gamma rays in opposite directions, which are detected by the PET scanner that surrounds the patient. In this way, the apparatus produces a series of plates across the body that combine to form a 3-D image.

The PET scanner can follow biological processes such as metabolism, neuronal transmission and tumor growth, so it is highly suitable for studying the behavior of medicines in the body, as well as for researching brain activity and for obtaining accurate 3-D images of tumors. In the future, antiprotons will also be used, instead of protons, in radiotherapy against tumors. This will allow to reduce substantially the level of radiation to the patient due to the greater destructive power of antiprotons when reaching tumor cells.

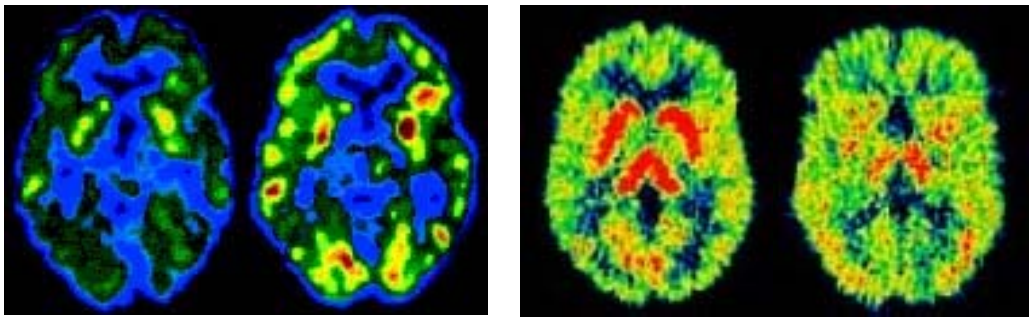

**Fig. 9.** On the left, one can see PET scans of the brain of an alcoholic patient 10 days (left) and 30 days (right) after starting the abstinence cure. On the right, PET scans of the brain of a healthy person (left) and a person with severe depression (right), revealing low levels of the neurotransmitter serotonin. *Credit*  Mac Gill University, Canada.

Regarding the technological applications of antimatter, the radiation with positrons - and in some cases also with antimuons - is widely used to study materials with important technological properties, such as ferroelectricity, superconductivity and magnetoresistance, as well as semiconductor materials used in solar panels and other electronic devices. These radiations are employed as well to obtain images of the inside of some materials for various purposes, from the analysis of processes of industrial interest (nuclear waste, oil flow...), up to finding structural defects in crystals or probing sub-nanometer scales in biopolymers.

Recommended Reading

B. Gato-Rivera, *Antimatter. What It Is and Why It´s Important in Physics and Everyday Life*. Switzerland, Springer Nature, 2021.

J. A. Caballero Carretero, *Dirac. La Antimateria*. Barcelona, RBA (Spanish), 2013.

N. Hall, *Antimatter. A review of its role in the universe and its applications*. Booklet from the IOP (Institute of Physics), UK, 2013.

F. Close, *Antimatter*. Oxford University Press, 2009.

G. Fraser, *Antimatter. The Ultimate Mirror*. Cambridge University Press, 2000.

S. Weinberg, *The First Three Minutes: A Modern View of the Origin of the Universe*. New York, Basic Books, 1$^{st}$ ed. 1977, 2$^{nd}$ ed. 1993.

P. A. M. Dirac, *The quantum theory of the electron*. Proceedings of the Royal Society A 117, 610, 1928.

P. A. M. Dirac, *Quantized singularities in the electromagnetic field*. Proceedings of the Royal Society A 133, 60, 1931.